\documentclass[journal]{IEEEtran}
\usepackage{amsmath,amsfonts}
\usepackage{algorithmic}
\usepackage{array}
\usepackage[caption=false,font=normalsize,labelfont=sf,textfont=sf]{subfig}
\usepackage{textcomp}
\usepackage{stfloats}
\usepackage{url}
\usepackage{verbatim}
\usepackage{graphicx}
\usepackage{xcolor}
\usepackage{tikz}
\usepackage{bbm}
\usetikzlibrary{arrows,shapes,chains,matrix,positioning,scopes,patterns,calc,
decorations.markings,
decorations.pathmorphing,
}
\usepackage{pgfplots}
\usepackage{ctable}
\pgfplotsset{compat=1.3}
\usepgflibrary{shapes}
\usepackage[nolist]{acronym}
\def\BibTeX{{\rm B\kern-.05em{\sc i\kern-.025em b}\kern-.08em
    T\kern-.1667em\lower.7ex\hbox{E}\kern-.125emX}}
\usepackage{balance}
\usepackage{soul}

\newcommand {\Define} {\stackrel {\Delta} {=}  } 

\begin{document}
\title{Zak-OTFS Based Coded Random Access for \\ Uplink mMTC
\thanks{The Duke team is supported in part by the National Science Foundation under grants 2342690 and 2148212, and is supported in part by funds from federal agency and industry partners as specified in the Resilient \& Intelligent NextG Systems (RINGS) program. The Duke team is also supported in part by the Air Force Office of Scientific research under grants FA 8750-20-2-0504 and FA 9550-23-1-0249.
Texas A\&M team is supported in part by the National Science Foundation under grant CNS-2148354.
University of Bologna team is supported by the European Union - Next Generation EU under the Italian National Recovery and Resilience Plan (NRRP), partnership on ``Telecommunications of the Future'' (PE00000001 - program “RESTART”). \\
\par \textbf{This work may be submitted to the IEEE/Springer for possible publication. Copyright
may be transferred without notice, after which this version may no longer be
accessible.}}
}
\author{
\IEEEauthorblockN{Alessandro Mirri\IEEEauthorrefmark{1}\textsuperscript{\textsection}, Venkatesh Khammammetti\IEEEauthorrefmark{2}\textsuperscript{\textsection}, Beyza Dabak\IEEEauthorrefmark{2}, Enrico Paolini\IEEEauthorrefmark{1}, \\ Krishna Narayanan\IEEEauthorrefmark{3}, and Robert Calderbank\IEEEauthorrefmark{2}}

\IEEEauthorblockA{\IEEEauthorrefmark{1}\textit{Department of Electrical, Electronic, and Information Engineering,
University of Bologna}, Bologna, Italy \\ \IEEEauthorrefmark{2}\textit{Electrical and Computer Engineering Department, Duke University}, Durham, NC, USA \\ \IEEEauthorrefmark{2}\textit{Department of Electrical and Computer Engineering, Texas A\&M
University}, College Station, TX, USA \\  alessandro.mirri7@unibo.it, venkatesh.khammammetti@duke.edu, beyza.dabak@duke.edu, e.paolini@unibo.it, krn@tamu.edu, and robert.calderbank@duke.edu\vspace{-1.2em}
   }
   }


\maketitle
\begingroup\renewcommand\thefootnote{\textsection}
\footnotetext{These two authors contributed equally to this work.}
\endgroup

\begin{acronym}
\small
\acro{ACK}{acknowledgement}
\acro{ARQ}{automatic retransmission request}
\acro{AWGN}{additive white Gaussian noise}
\acro{BCH}{Bose–Chaudhuri–Hocquenghem}
\acro{BN}{burst node}
\acro{BS}{base station}
\acro{CDF}{cumulative distribution function}
\acro{CP}{cyclic prefix}
\acro{CRA}{coded random access}
\acro{CRC}{cyclic redundancy check}
\acro{CRDSA}{contention resolution diversity slotted ALOHA}
\acro{CSA}{coded slotted ALOHA}
\acro{DD}{delay-Doppler}
\acro{DE}{density evolution}
\acro{DLD}{double-latency deadlines}
\acro{DSA}{diversity slotted ALOHA}
\acro{eMBB}{enhanced mobile broad-band}
\acro{FER}{frame error rate}
\acro{HLL}{high-low-low}
\acro{ICI}{inter carrier interference}
\acro{IFSC}{intra-frame spatial coupling}
\acro{i.i.d.}{independent and identically distributed}
\acro{IoT}{Internet of Things}
\acro{I/O}{input-output}
\acro{IRSA}{irregular repetition slotted ALOHA}
\acro{ISI}{inter-symbol interference}
\acro{KPI}{key performance indicator}
\acro{LDPC}{low-density parity-check}
\acro{LOS}{line of sight}
\acro{MAC}{medium access control}
\acro{MIMO}{multiple-input multiple-output}
\acro{ML}{maximum likelihood}
\acro{MMA}{massive multiple access}
\acro{MA}{multiple access}
\acro{MMSE}{minimum mean squared error}
\acro{mMTC}{massive machine-type communications}
\acro{MPR}{multi-packet reception}
\acro{MRC}{maximal ratio combining}
\acro{MTC}{machine-type communications}
\acro{NOMA}{non-orthogonal multiple access}
\acro{OFDM}{orthogonal frequency-division multiplexing}
\acro{PAB}{payload aided based}
\acro{PB}{power balance}
\acro{PDF}{probability density function}
\acro{PDMA}{power domain multiple access}
\acro{PGF}{probability generating function}
\acro{PHY}{physical}
\acro{PLR}{packet loss rate}
\acro{PMF}{probability mass function}
\acro{PRCE}{perfect replica channel estimation}
\acro{PU}{power unbalance}
\acro{QAM}{quadrature amplitude modulation}
\acro{QPSK}{quadrature phase-shift keying}
\acro{RedCap}{Reduced Capacity}
\acro{RF}{radio-frequency}
\acro{SA}{slotted ALOHA}
\acro{SC}{single-carrier}
\acro{SSC}{spaced spatial coupling}
\acro{SIC}{successive interference cancellation}
\acro{SIS}{successive interference subtraction}
\acro{SLD}{single-latency deadline}
\acro{SN}{sum node}
\acro{SNB}{squared norm based}
\acro{SNR}{signal-to-noise ratio}
\acro{TF}{time-frequency}
\acro{URLLC}{ultra-reliable and low-latency communication}
\acro{Zotfs}{Zak-OTFS}
\acro{Zak-OTFS}{Zak-orthogonal time frequency space}
\end{acronym}

\begin{abstract}
This paper proposes a grant-free coded random access (CRA) scheme for uplink massive machine-type communications (mMTC), based on Zak-orthogonal time frequency space (Zak-OTFS) modulation in the delay-Doppler domain.
The scheme is tailored for doubly selective wireless channels, where conventional orthogonal frequency-division multiplexing (OFDM)-based CRA suffers from unreliable inter-slot channel prediction due to time-frequency variability.
By exploiting the predictable nature of Zak-OTFS, the proposed approach enables accurate channel estimation across slots, facilitating reliable successive interference cancellation across user packet replicas.
A fair comparison with an OFDM-based CRA baseline shows that the proposed scheme achieves significantly lower packet loss rates under high mobility and user density. Extensive simulations over the standardized Veh-A channel confirm the robustness and scalability of Zak-OTFS-based CRA, supporting its applicability to future mMTC deployments.
\end{abstract}


\section{Introduction}
\label{sec:Intro}
\Ac{mMTC} is a key use case in 5G and beyond. 
It is characterized by a massive number of devices that sporadically and unpredictably transmit short data packets to a central \ac{BS}, typically under stringent constraints on energy consumption and signaling overhead, and often mild latency and reliability constraints~\cite{Bockelmann2016:Massive,BocPraWun:18, Mahmood2020:white}.
To address this challenge of \ac{MMA} \cite{Wu2020:Massive,Chen2020:massive}, grant-free random access schemes \cite{casini2007:contention,liva2011:irsa,paolini2015:csa} have gained increasing attention, as they allow devices to transmit autonomously, whenever new data is available, without requiring explicit scheduling by the \ac{BS}. 
While this uncoordinated approach reduces latency and protocol complexity, it also increases the likelihood of packet collisions among simultaneously transmitting users, potentially leading to information loss.
As a result, achieving efficient and reliable communication under these constraints necessitates a carefully integrated design of the \ac{MAC} and \ac{PHY} layers.
\Ac{SIC}-based \ac{CRA} \cite{Berioli2016:Modern} has emerged as an effective and pragmatic integrated \ac{MAC}/\ac{PHY} approach for the design of grant-free random access.

Seminal works on integrated \ac{MAC}/\ac{PHY} designs were tailored to the collision channel
\cite{casini2007:contention,liva2011:irsa}, which corresponds to the packet erasure channel at the \ac{PHY} layer; hence, they did not account for a power constraint.
In \cite{Polyanskiy2017:Perspective}, Polyanskiy proposed a theoretical framework to model grant-free random access for the power-constrained additive white Gaussian noise (AWGN) channel, which is now popularly referred to as unsourced random access. 
He also derived an achievable probability of error for random coding as a function of \ac{SNR}, which has served as a benchmark for grant-free random access coding schemes. 
Polyanksiy's results demonstrate that the performance of naive random access schemes such as \ac{SA} \cite{roberts1975aloha} is significantly far away from the random coding bound. 
Later works have shown that \ac{CRA}-based schemes can be optimized for the AWGN channel and that they offer significant performance improvement over naive \ac{SA} schemes \cite{vem2019:unsourced,Tralli2024:IrsaAWGN}, particularly for a large number of users. 
Subsequent works have extended \ac{CRA}-based schemes to flat Rayleigh fading channels \cite{clazzer2017:irregular,liva2019:on-off}.

Despite these advances, existing \ac{CRA} and SIC-based schemes are typically designed for simplified channel models (AWGN or flat Rayleigh fading) that fail to capture the time and frequency selectivity of realistic wireless channels. 
In particular, when the channel exhibits both significant delay and Doppler spreads, the effective channel response varies rapidly within a time-frequency resource block (slot). 
This variability poses a major challenge for \ac{SIC}-based architectures, which rely on accurate estimation of the \ac{I/O} relationship over the entire transmission slot. 
Modulation schemes such as \ac{OFDM} and \ac{SC} modulation are ill-suited for this setting: OFDM offers simplified equalization; however, it requires prior knowledge of the channel model for effective \ac{I/O} acquisition, and \ac{SC} suffers from requiring a significant number of pilots for estimating doubly-spread channels. 

In this paper, we engineer a \ac{CRA} scheme based on \ac{Zak-OTFS} modulation \cite{Saif_Bits1} and show that our proposed scheme is significantly more effective than OFDM-based schemes for enabling \ac{SIC}. In Zak-OTFS modulation, the carrier waveform is a pulse in the \ac{DD} domain, formally a quasi-periodic localized function with specific periods along delay and Doppler. 
When the channel delay spread is less than the delay period, and the channel Doppler spread is less than the Doppler period, the response to a single Zak-OTFS carrier provides an image of the scattering environment, which is used to predict the effective channel.
Since the scattering environment changes slowly and predictably, our approach enables accurate prediction of the channel and hence, it makes \ac{SIC} viable.
While Zak-OTFS introduces complexity in equalization due to \ac{ISI}, it offers an important advantage over OFDM: the acquisition of the \ac{I/O} relationship can be achieved in a model-agnostic manner. We demonstrate that \ac{CRA} and \ac{SIC}, when integrated with Zak-OTFS, significantly outperforms OFDM-based schemes in doubly selective channels, thereby enabling efficient and reliable grant-free random access for \ac{mMTC}.

The main contributions of this paper are summarized as follows:
\begin{itemize}
    \item We propose a novel grant-free \ac{CRA} scheme designed directly in the \ac{DD} domain, leveraging \ac{Zak-OTFS} modulation to address the challenges of random access over doubly selective wireless channels.
    Unlike conventional approaches that operate in the \ac{TF} domain, the proposed scheme exploits the predictable structure of the Zak-OTFS modulation in the \ac{DD} domain to enable replica decoding and effective \ac{SIC} across slots.
    \item We analyze the impact of pulse shaping filter design on Zak-OTFS performance, comparing Gaussian and sinc filters. This design choice enables balanced trade-offs in slot-level processing across different system performance metrics, depending on the specific system settings.
    \item We verify that under high-mobility scenarios, the performance of the \ac{OFDM}-based scheme degrades significantly due to its reliance on inter-slot channel prediction, which compromises the effectiveness of \ac{SIC}. In contrast, the proposed \ac{Zak-OTFS}-based approach maintains robust and reliable \ac{SIC} processing, enabled by its near-invariant \ac{I/O} relationship across the entire frame.
    \item We perform extensive simulations under realistic Veh-A channel conditions, confirming that the proposed Zak-OTFS-based \ac{CRA} scheme consistently achieves lower \ac{PLR} than the \ac{OFDM} baseline, particularly in high-mobility and high user density scenarios.
\end{itemize}
These results highlight the practical advantages of designing \ac{CRA} schemes in the \ac{DD} domain and establish \ac{Zak-OTFS} as a viable modulation framework for scalable and reliable uplink \ac{mMTC}.

\section{Preliminaries and Background}
\label{sec:Preliminaries}
The \ac{MAC} layer in a wireless network manages contention resolution, resource selection, and interference cancellation, while the \ac{PHY} layer governs how information is encoded, modulated, and transmitted over the wireless channel.
Meeting the stringent performance requirements of \ac{mMTC}, particularly in terms of scalability under latency and reliability constraints,
requires a joint and well-coordinated \ac{MAC}/\ac{PHY} design.
This section is therefore divided into two parts: the first subsection reviews key principles and recent developments in random access protocols at the \ac{MAC} layer, with particular emphasis on contention resolution and the role of \ac{SIC}; the second provides the necessary \ac{PHY} layer background on Zak-OTFS modulation, which serves as the foundation of the \ac{PHY} layer model adopted in this work.

\subsection{MAC Layer: Random Access and SIC}
\label{subsec:BackgroundMAC}
%
\begin{figure*}[t]
	\centering
        \includegraphics[width=1.0\textwidth]{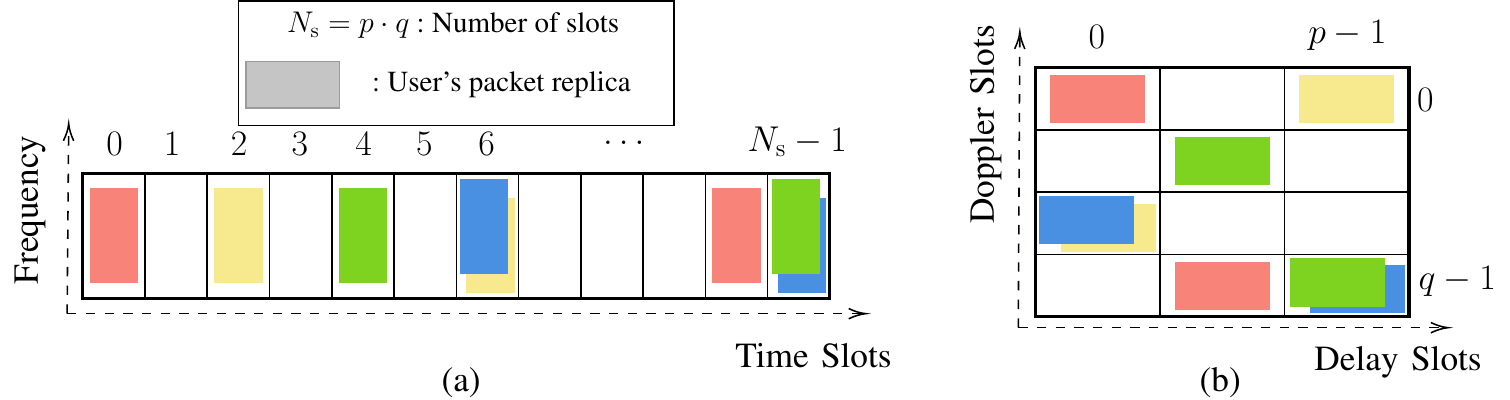}
	\caption{Comparison between (a) conventional coded random access in the time-frequency domain and (b) proposed Zak-OTFS-based coded random access in the delay-Doppler domain. In both cases, colored rectangles represent user-specific packet replicas transmitted in randomly selected slots within the corresponding resource grid.}
    \label{fig:Scenario_OTFSvsOFDM_Frame}
\end{figure*}
Random access protocols have historically played a central role in enabling uncoordinated communication in shared wireless networks.
A foundational example is \ac{SA}, where time is divided into slots and active users independently attempt transmission of a single information packet without any coordination.
The analysis typically relies on the \textit{collision channel model}, which  assumes that a packet is successfully received only if it is the sole transmission in a given slot; otherwise, all colliding packets are lost.
It is further assumed that collisions are always detectable at the \ac{BS} and that, at the end of each slot, a feedback message is broadcast to all users.
This feedback enables each user to determine whether its packet was successfully received or if a retransmission is required in subsequent time slots.
While simple and fully decentralized, this one-shot transmission strategy suffers from a high probability of collisions, especially as the number of active users grows, leading to poor throughput and reduced reliability.

To overcome these limitations and improve the system performance, more advanced random access schemes introduced temporal redundancy, where each user transmits multiple replicas of the same packet in randomly selected time slots.
This strategy exploits slot diversity, increasing the probability that at least one replica avoids a collision and is successfully received.
An early example of this idea is \ac{DSA}~\cite{choudhury1983diversity}, which demonstrated improved reliability over traditional \ac{SA} by reducing the probability of complete packet loss due to collisions.

A major breakthrough came with the introduction of \ac{SIC}, which enables the receiver to iteratively decode packets and subtract their reconstructed contributions from the received signal. 
This process allows for the resolution of packet collisions that would otherwise lead to data loss.
This principle forms the foundation of modern \ac{CRA} protocols such as \ac{CRDSA}~\cite{casini2007:contention} and \ac{IRSA}~\cite{liva2011:irsa}, which model the access process as a sparse bipartite graph and adopt belief-propagation-inspired decoding techniques originally developed for erasure codes.
In these schemes, each user transmits a predefined \cite{casini2007:contention} or randomly selected \cite{liva2011:irsa} number of replicas in independently chosen time slots within a contention window, commonly referred to as a frame.
This structure is illustrated in Fig.~\ref{fig:Scenario_OTFSvsOFDM_Frame}(a), where the frame is divided into $N_\mathrm{s}$ discrete time slots, each representing a potential transmission opportunity for a packet replica.
The \ac{BS} then leverages the graph structure of the collisions to iteratively apply \ac{SIC}, progressively removing interference and recovering additional packets.

For \ac{CRA} schemes to operate effectively and support reliable \ac{SIC}, the conventional assumptions of the collision channel model are typically extended to include ideal interference subtraction across slots.
Under this model, perfect \ac{SIC} is assumed, meaning that once a packet is successfully decoded, its contribution can be perfectly removed from the received signal in all time slots where its replicas were transmitted, regardless of the actual channel conditions.

However, while this assumption enables tractable analysis and protocol design, it neglects critical \ac{PHY} layer impairments encountered in practical wireless deployments.
In reality, the wireless channel coefficients for the same user may vary across the frame, depending on the relationship between the channel coherence time and the frame duration. 
This effect is particularly pronounced in high-mobility environments, such as industrial or vehicular networks, where Doppler shifts and delay spreads significantly reduce the channel coherence time, violating the assumption of invariant channels across replicas.
As a result, replicas of the same packet are likely to experience different channel conditions, causing residual interference and impairing the effectiveness of \ac{SIC}.

To overcome these challenges, the channel coefficients must be accurately estimated at the \ac{BS} not only in the slots where a packet is successfully decoded, but also in all slots where interference needs to be removed through \ac{SIC}.
Accordingly, recent research has explored more advanced \ac{BS} architectures and sophisticated decoding algorithms that go beyond the idealized \ac{SIC} assumption, enabling more effective interference cancellation under realistic channel conditions.
In particular, equipping the \ac{BS} with a massive \ac{MIMO} architecture introduces spatial degrees of freedom that can be effectively exploited to enhance collision resolution and boost the performance of \ac{SIC}-based decoding.
Recent studies, such as~\cite{Valentini2022:Massive, Val23:SICAlgo, Valentini2024:Feedback, Mirri2025:Latency}, have demonstrated that spatial processing in massive \ac{MIMO} systems significantly improves the effectiveness of \ac{SIC}, even under rapidly varying channel conditions that change on a slot-by-slot basis~\cite{Bjo2017:MIMObook, Sor2018:coded}.
However, as noted, these approaches concentrate the majority of the system burden at the \ac{BS}, both in terms of architecture and receiver-side signal processing, without directly addressing potential improvements on the transmission side, such as the selection of a more suitable \ac{PHY} layer waveform.

In this work, we adopt a complementary perspective by shifting the focus from conventional \ac{TF} domain \ac{PHY} processing to a \ac{DD} representation, leveraging the Zak-OTFS waveform as the modulation framework.
Rather than relying on traditional \ac{CRA} schemes in the \ac{TF} domain, as depicted in Fig.~\ref{fig:Scenario_OTFSvsOFDM_Frame}(a), we propose a novel \ac{CRA} strategy designed in the \ac{DD} domain, as illustrated in Fig.~\ref{fig:Scenario_OTFSvsOFDM_Frame}(b) and further detailed in Section~\ref{sec:ProposedUplinkMA}.
This shift is motivated by the observation that wireless channels, often highly dynamic and unpredictable in the \ac{TF} domain, tend to exhibit a more structured and nearly static behavior in the \ac{DD} domain.
In particular, the Zak-OTFS representation makes the channel approximately invariant over the entire duration of \ac{DD} frame \cite{Saif_Bits1,Saif_Bits2}, even in high-mobility scenarios.
As a result, assumptions such as ideal \ac{SIC}, typically violated in \ac{TF} settings, become significantly more realistic in the \ac{DD} domain, enabling more robust and efficient grant-free random access in challenging \ac{mMTC} environments.

\subsection{PHY Layer: Zak-OTFS}
\label{subsec:BackgroundPHY}
Zak-OTFS modulation is a variant of \ac{DD} modulation  \cite{SaifBook, Survey_paper, hong2022delay} that multiplexes information symbols in the \ac{DD} domain. In \ac{DD} modulation, the delay period $\tau_p$ and the Doppler period $\nu_p$ are divided into $M$ and $N$ uniform intervals, respectively, such that the delay resolution is $\frac{\tau_p}{M}$ and the Doppler resolution is $\frac{\nu_p}{N}$.
Also, we assume the wireless channel is doubly spread so that  $\tau_p \nu_p = 1$. 
\par In Zak-OTFS modulation, the information symbols $x[k,l]$, with indices $k = 0,1, \cdots, M-1$ and $l = 0,1, \cdots, N-1$, are arranged on an $M \times N$ grid and mapped to the \ac{DD} lattice $ \Lambda_{\mathrm{dd}} \Define  \{ (k\tau_p/M, l\nu_p/N) \, | \, k,l \in {\mathbb Z} \}$. The quasi-periodic \ac{DD} domain signal\footnote{Note that the time Zak transform exists only for quasi-periodic signals in the \ac{DD} domain. Consequently, only quasi-periodic \ac{DD} domain signals admit a valid representation in the time domain.} $x_{\text{dd}}[k+nM, l+mN]$ is generated such that
\begin{align}
\label{input_qp}
    x_{\text{dd}}[k+nM, l+mN] =
    \begin{cases}
        x[k,l], & \text{if } n = m = 0, \\
        x[k,l] e^{j2\pi n l / N}, & \text{otherwise,}
    \end{cases}
\end{align}
where $n,m \in \mathbb{Z}$. This discrete \ac{DD} domain signal is then mapped to the continuous domain as
\begin{align} 
\label{cont_dd}
    x_{\text{dd}}(\tau,\nu) = \sum_{k,l \in \mathbb{Z}} x_{\text{dd}}[k,l] \delta\left(\tau - \frac{k\tau_p}{M}\right) \delta\left(\nu - \frac{l\nu_p}{N}\right),
\end{align}
where $\tau, \nu \in \mathbb{R}$ denote the continuous delay and Doppler variables, respectively. The signal in (\ref{cont_dd}) is then passed through a transmit shaping filter $w_{\text{tx}}(\tau,\nu)$ resulting in
\begin{align}
\label{tx_fil_dd}
    x^{\text{tx}}_{\text{dd}}(\tau,\nu) = w_{\text{tx}}(\tau,\nu) *_\sigma x_{\text{dd}}(\tau,\nu),
\end{align}
 where $*_\sigma$ denotes the twisted convolution\footnote{$a(\tau,\nu) *_\sigma b(\tau,\nu) = \iint a(\tau',\nu')b(\tau-\tau',\nu-\nu')e^{j2\pi\nu'(\tau-\tau')} \, d\nu' d\tau'$.} operation. Note that the choice of transmit filter $w_{\text{tx}}(\tau,\nu)$ is a design degree of freedom that can be used to balance different measures of system performance \cite{Closed_form_OTFS,Gauss_Sinc}.

\par The doubly spread wireless channel is modeled through its \ac{DD} impulse response $h_{\text{phy}}(\tau,\nu) = \sum_{\ell=1}^{L}h_\ell \delta(\tau-\tau_\ell)\delta(\nu-\nu_\ell)$, where $h_\ell,\tau_\ell$ and $\nu_\ell$ are the channel gain, delay spread and Doppler spread of the $\ell$-th path, respectively, and $L$ is the total number of paths.
The transmit signal $x^{\text{tx}}_{\text{dd}}(\tau,\nu)$ propagates through this \ac{DD} domain channel, resulting in the received signal
\begin{align} 
\label{rx_cont}
    y_{\text{dd}}(\tau,\nu) = h_{\text{phy}}(\tau,\nu)  *_\sigma x^{\text{tx}}_{\text{dd}}(\tau,\nu) + n(\tau,\nu),
\end{align}
where $n(\tau,\nu)$ represents additive noise in the \ac{DD} domain, corresponding to a time-domain noise with power spectral density $N_0$. 
\par In Zak-OTFS demodulation, the received \ac{DD} domain signal $y_{\text{dd}}(\tau,\nu)$ is first filtered using the receive pulse $w_{\text{rx}}(\tau,\nu)$, yielding
\begin{align}
\label{rx_dd}
    y^{\text{rx}}_{\text{dd}}(\tau,\nu) = w_{\text{rx}}(\tau,\nu) *_\sigma y_{\text{dd}}(\tau,\nu).
\end{align}
 The output $y^{\text{rx}}_{\text{dd}}(\tau,\nu)$ is then sampled at $\tau=k\tau_p/M$ and $\nu=l\nu_p/N$, resulting in the discrete received symbols
\begin{align}
\label{rxsymbols}
    y[k,l] &= y^{\text{rx}}_{\text{dd}}\left(\tau = \frac{k\tau_p}{M}, \nu = \frac{l\nu_p}{N}\right) \nonumber \\
    &= \sum_{k^{\prime},l^{\prime} \in \mathbb{Z}} h_{\text{eff}}[k-k^{\prime},l-l^{\prime}] x_{\text{dd}}[k^{\prime},l^{\prime}] e^{j2\pi \left(\frac{(l - l^{\prime})k^{\prime}}{M}\right)},
\end{align}
where $k = 0,1,\cdots, M-1$ and $l = 0,1,\cdots,N-1$.
The discrete effective channel response $h_{\text{eff}}[k,l]$ is obtained by sampling the continuous effective channel
\begin{align}
\label{h_eff_exp}
    h_{\text{eff}}(\tau,\nu) = w_{\text{rx}}(\tau,\nu) *_\sigma h_{\text{phy}}(\tau,\nu) *_\sigma w_{\text{tx}}(\tau,\nu).
\end{align}
The effective channel $h_{\text{eff}}[k, l]$ can be accurately predicted in Zak-OTFS, provided that the channel delay spread is less than the delay period and the channel Doppler spread is less than the Doppler period (this is the \textit{crystalline condition} introduced in \cite{Saif_Bits1, Saif_Bits2}). 
\par In the next section, we present the proposed uplink \ac{MMA} scheme for time-varying channels, integrating both the \ac{MAC} and \ac{PHY} layer aspects discussed above.

\section{Proposed Uplink MMA Scheme}
\label{sec:ProposedUplinkMA}
\subsection{OTFS-Based Framing and Random Access Protocol}
\label{OTFS_frame_RA}
\begin{figure}[]
    \centering
    \includegraphics[width=0.48\textwidth]{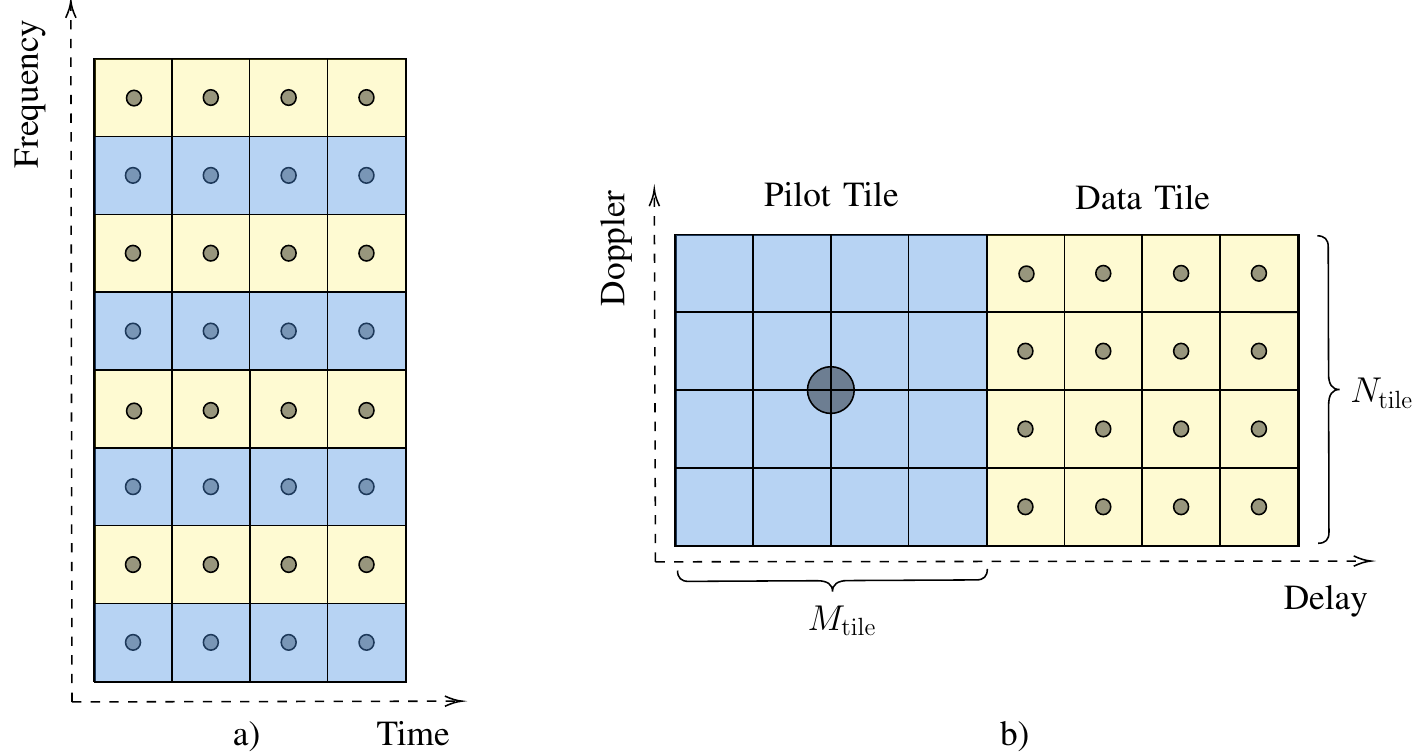}
    \caption{ 
    Structure of a single slot in the coded random access scheme, representing one slot within the overall frame illustrated in Fig.~\ref{fig:Scenario_OTFSvsOFDM_Frame}.
        \textbf{(a)} Time-frequency domain representation of the slot, where pilot and data symbols are spread across time and frequency resources, following a structure typical of OFDM-based transmissions.
        \textbf{(b)} Delay-Doppler domain representation of the slot, comprising two tiles: a pilot tile (blue) and a data tile (yellow), each of size $M_{\text{tile}} \times N_{\text{tile}}$.
        The pilot tile includes a single point pilot symbol centered in the grid (large circle), while the data tile carries modulated data symbols (small circles).
        }
    \label{fig:Scenario_OTFSvsOFDM_SingleSlot}
\end{figure}
We consider a \ac{CRA} scheme operating in the \ac{DD} domain based on Zak-OTFS modulation.
In the proposed setup, each user transmits a complete OTFS frame in the uplink.
The frame is divided into multiple slots, with each slot occupying $M_\mathrm{slot}$ and $N_\mathrm{slot}$ resources along the delay and Doppler axes, respectively. 
The full frame spans $M \times N = p \cdot M_{\text{slot}} \times q \cdot N_{\text{slot}}$ resources, where $p$ and $q$ denote the number of slots arranged along the delay and Doppler dimensions, as shown in Fig. \ref{fig:Scenario_OTFSvsOFDM_Frame}(b).
The fundamental unit of an OTFS slot is the \textit{tile}, defined as a region of size $M_{\text{tile}} \times N_{\text{tile}}$ in the \ac{DD} grid (see Fig.~\ref{fig:Scenario_OTFSvsOFDM_SingleSlot}(b)).
Each tile is allocated either to pilot or data (payload) transmission.
A single OTFS slot is formed by concatenating two adjacent tiles; i.e., one for pilot and one for data.
This concatenation can occur along either the delay or Doppler axis; for simplicity, we adopt concatenation along the delay axis, which yields slot dimensions of $M_{\text{slot}} \times N_{\text{slot}} = 2M_{\text{tile}} \times N_{\text{tile}}$.
While the choice of axis affects indexing, it does not impact system operation as long as the crystalline condition is satisfied.

In the proposed protocol, each active user independently selects $r$ non-overlapping OTFS slots with uniform probability within the frame for transmission.
Selecting a slot\footnote{In this paper, we use the term \textit{slot} to refer specifically to an OTFS slot.} entails choosing both a pilot tile and its associated data tile.
The user then transmits an identical replica of its packet in each selected slot, i.e., the same pilot symbol is placed in each pilot tile, and the same modulated data symbols are used in each corresponding data tile.
This replica consistency is crucial for enabling coherent \ac{SIC} at the receiver, as it ensures that all copies carry identical information.
Overall, each active user occupies $r \cdot M_{\text{slot}} \cdot N_{\text{slot}}$ \ac{DD} resources, distributed across $r$ slots over the entire Zak-OTFS frame, which spans a bandwidth $B = \nu_p \cdot M$ and duration $T = \tau_p \cdot N$.

\subsection{Signal Processing at the Transmitter}
\label{subsec:TxProcessing}
Consider an uplink \ac{MMA} system where $K_\mathrm{a}$ users aim to communicate simultaneously with the \ac{BS}. Let $x_{_{k_\mathrm{a}}}[k,l]$, where $k = 0,1, \cdots, M-1$ and $l = 0,1, \cdots, N-1$, denote the information transmitted by the $k_\mathrm{a}$-th user over the entire DD frame.
Let $s_{_{k_\mathrm{a}}}[\tilde{k},\tilde{l}]$, with $\tilde{k} = 0,1,\cdots, 2M_{\text{tile}}-1$ and $\tilde{l} = 0,1, \cdots, N_{\text{tile}}-1$, represent the symbols transmitted by the $k_\mathrm{a}$-th user within a single slot\footnote{We refer to this sequence of symbols as the user packet.} (see Section~\ref{OTFS_frame_RA} for details).
For $\tilde{k} = 0,1,\cdots, M_{\text{tile}}-1$ and $\tilde{l} = 0,1, \cdots, N_{\text{tile}}-1$, $s_{_{k_\mathrm{a}}}[\tilde{k},\tilde{l}]$ corresponds the pilot tile. 
In this region, the user transmits a single pilot symbol at the center position $(k_p,l_p) = (M_{\text{tile}}/2, N_{\text{tile}}/2)$ with all other symbols set to zero\footnote{This configuration resembles a point pilot surrounded by a guard band, which serves to prevent interference (between the pilot and data regions) caused by channel spreading.}. For $\tilde{k} = M_{\text{tile}},M_{\text{tile}}+1,\cdots, 2M_{\text{tile}}-1$ and $\tilde{l} = 0,1, \cdots, N_{\text{tile}}-1$, $s_{_{k_\mathrm{a}}}[\tilde{k},\tilde{l}]$ defines the data tile, where the user transmits modulated information symbols. 
As illustrated in Fig.~\ref{fig:Scenario_OTFSvsOFDM_Frame}(b), the \ac{DD} frame is partitioned into $N_\mathrm{s} = p\cdot q$ slots. These slots are indexed sequentially from $0$ to $N_\mathrm{s}-1$ and arranged into a $p \times q$ slot index matrix following a row-wise order. This matrix representation facilitates efficient indexing and systematic access to slot locations during transmission and reception. 
Let $a \in \{0,1, \cdots, N_\mathrm{s}-1\}$ denote a linear slot index.
Its corresponding matrix indices $(i,j)$ are given by
\begin{equation}
    \begin{aligned}
        i &=  \left\lfloor \frac{a}{q} \right\rfloor, \,\ i \in \{0,1, \cdots, p-1\},  \\
        j &= \left(a\right)_{q}, \,\ j \in \{0,1, \cdots, q-1\}, 
\end{aligned}
\end{equation}
where $(\cdot)_q$ denotes the modulus-$q$ operation. Let $\mathcal{A} = \{a_0, a_1, \dots, a_{r-1}\} \subset \{0, 1, \dots, N_s - 1\}$ be the set of $r$ selected slot indices. Each slot index $a_z \in \mathcal{A}$ maps to coordinates $(i_z,j_z)$ computed as
\begin{equation}
    i_z = \left\lfloor \frac{a_z}{q} \right\rfloor,\quad 
    j_z = \left(a_z\right)_{q}, \nonumber
\end{equation}
where $z = 0,1, \cdots, r-1$. The $k_\mathrm{a}$-th user packet $s_{_{k_\mathrm{a}}}[\tilde{k},\tilde{l}]$ is replicated across the $r$ selected slots to effectively distribute the transmission across multiple \ac{DD} resources. 
Therefore, the corresponding transmit signal over the \ac{DD} frame, $x_{_{k_\mathrm{a}}}[k,l]$,
is expressed as
\begin{align}
\label{dd_tx_sig}
    x_{_{k_a}}[k,l] &= \sum_{z=0}^{r-1} s_{_{k_\mathrm{a}}}[k - 2M_{\text{tile}} i_z,\; l - N_{\text{tile}} j_z] 
 \nonumber \\ 
& \cdot \mathbbm{1}_{[2M_{\text{tile}} i_z,\; 2M_{\text{tile}}(i_z+1))}(k) \cdot 
\mathbbm{1}_{[N_{\text{tile}} j_z,\; N_{\text{tile}}(j_z+1))}(l),
\end{align}
where $\mathbbm{1}_{[c, d)}(x)$ is the indicator function:
\[
    \mathbbm{1}_{[c, d)}(x) = 
    \begin{cases}
        1 & \text{if } x \in [c, d) \\
        0 & \text{otherwise.}
    \end{cases}
\]
The function $s_{_{k_\mathrm{a}}}[m', n']$ is defined only for \( 0 \leq m' < 2M_{\text{tile}} \) and \( 0 \leq n' < N_{\text{tile}} \), and the values $(i_z,j_z)$ ensure that $s_{_{k_\mathrm{a}}}[m',n']$ is mapped to the desired slot. The replication strategy in~\eqref{dd_tx_sig} ensures that each user’s transmission spans the entire OTFS frame, enhancing diversity and robustness against channel impairments such as delay and Doppler spreads. 
Following the Zak-OTFS signal processing described in Section \ref{subsec:BackgroundPHY}, the user's \ac{DD} domain signal $x_{_{k_\mathrm{a}}}[k,l]$ is processed via~\eqref{input_qp}, \eqref{cont_dd} and \eqref{tx_fil_dd} to produce the continuous transmit signal $x_{\text{dd},k_\mathrm{a}}^{\text{tx}}(\tau,\nu)$.
 In this paper, we employ two \ac{DD} domain transmit filters $w_{\text{tx}}(\tau,\nu)$, as detailed in Table \ref{DD_filters_design}\footnote{We do not consider the Gaussian-Sinc filter~\cite{Gauss_Sinc} due to its mathematical complexity. Numerical results show that, for the proposed \ac{MMA} scheme, its performance is close to the Sinc filter and not better than the Gaussian filter. Similarly, the RRC filter is omitted due to the lack of a tractable closed-form expression for $h_{\text{eff}}[k,l]$ \cite{Closed_form_OTFS}.}.
\begin{table}[!t]
\caption{DD Filters~\cite{Gauss_filter}.}
\centering
\label{DD_filters_design}
\begin{tabular}{|c!{\vrule width 0.1em}c|c|}
\hline
DD Filters & $w_{tx}(\tau,\nu) = w_1(\tau) w_2(\nu)$   \\
\specialrule{.1em}{.05em}{.05em} 
Sinc Filter &  
$\begin{aligned} & w_1(\tau) \Define\sqrt{B}\operatorname{sinc}(B\tau) \\ & w_2(\nu) \Define\sqrt{T}\operatorname{sinc}(T\nu)\end{aligned}$
\\
\hline
Gaussian Filter  &  $\begin{aligned} & w_1(\tau) \Define\left(\frac{2 \alpha_\tau B^2}{\pi}\right)^{\frac{1}{4}} e^{-\alpha_\tau B^2 \tau^2} \\ & w_2(\nu) \Define\left(\frac{2 \alpha_\nu T^2}{\pi}\right)^{\frac{1}{4}} e^{-\alpha_\nu T^2 \nu^2}\end{aligned}$   \\
\hline
\end{tabular}
\vspace{-1.3em}
\end{table}
At the receiver, we use a matched filter in the \ac{DD} domain, which is defined as~\cite{Gauss_filter} 
 \begin{align}
 \label{match_filter}
     w_{rx}(\tau,\nu) = e^{j2\pi\nu\tau}w_{tx}^{*}(-\tau,-\nu).
 \end{align}
Closed-form expressions for the effective channel $h_{\text{eff}}[k,l]$ under each filter choice can be obtained using Table~\ref{DD_filters_design}, (\ref{h_eff_exp}) and (\ref{match_filter}).
We refer the reader to~\cite{Closed_form_OTFS} for details.

\subsection{Slot-Level Signal Processing at the Receiver }
\label{subsec:OTFS_RxProcessing}
Assuming all active users are frame-synchronized, they transmit within a shared uplink OTFS frame. Consequently, the received \ac{DD} domain symbols at the \ac{BS}, denoted by $y[k^{\prime},l^{\prime}]$, for $k^{\prime} = 0,1, \cdots, M-1$ and $l^{\prime} = 0,1, \cdots, N-1$, consist of the superposition of the contributions from all $K_\mathrm{a}$ active users, i.e.,
 \begin{align}
     \label{received_uplink}
     y[k^{\prime},l^{\prime}] &= \sum_{k_\mathrm{a}=1}^{K_\mathrm{a}} \tilde{y}_{_{k_\mathrm{a}}}[k^{\prime},l^{\prime}] + n[k',l'];\nonumber \\
     \tilde{y}_{_{k_\mathrm{a}}}[k^{\prime},l^{\prime}] & \Define \sum_{k^{\prime\prime},l^{\prime\prime} \in \mathbb{Z}} h_{\text{eff},k_\mathrm{a}}[k^{\prime}-k^{\prime\prime},l^{\prime}-l^{\prime\prime}] x_{_{\text{dd},{k_\mathrm{a}}}}[k^{\prime\prime},l^{\prime\prime}] \nonumber \\ & \cdot e^{j2\pi \left(\frac{(l^{\prime} - l^{\prime\prime})k^{\prime\prime}}{M}\right)} + n[k',l'].
 \end{align}
 Here, $\tilde{y}_{_{k_\mathrm{a}}}[k^{\prime},l^{\prime}]$ denotes the received contribution from the $k_\mathrm{a}$-th user, where $h_{\mathrm{eff},k_\mathrm{a}}[k^{\prime},l^{\prime}]$ is the corresponding user-specific effective channel response, and $x_{_{\text{dd},{k_\mathrm{a}}}}[k^{\prime},l^{\prime}]$ is the quasi-periodic \ac{DD} domain signal transmitted by the $k_\mathrm{a}$-th user, for all $k^{\prime},l^{\prime} \in \mathbb{Z}$.
At the receiver, signal processing is performed on a per-slot basis\footnote{Per-slot decoding and \ac{DD} domain analysis have been conducted in \cite{Venky_1, Augustine_1, Venky_2}, primarily within the context of MC-OTFS/OTFS 1.0 under an orthogonal multiple access (OMA) framework. In contrast, this work addresses randomized transmissions in a grant-free setting.}, where the channel response is estimated from the pilot tile and then used to equalize the symbols in the associated data tile. 
Assuming that the repetition pattern (or preambles) of all active users are known, the \ac{BS} begins by identifying and processing the \textit{singleton slots}, i.e., slots in which no collision occurs and only a single user transmits its packet\footnote{Alternatively, the \ac{BS} may operate in a fully blind manner, without prior knowledge of users' repetition patterns. However, this approach typically incurs higher decoding complexity.}.

\par Using the knowledge of the preambles, the BS identifies that the $\widetilde{k}_\mathrm{a}$-th user's signal contribution, denoted as $\tilde{y}_{_{\widetilde{k}_\mathrm{a}}}[k^{\dagger},l^{\dagger}]$ for $ k^{\dagger} = 0,1, \cdots, 2M_{\text{tile}}-1$ and $l^{\dagger} = 0,1, \cdots, N_{\text{tile}}-1$, is received within a singleton slot indexed by $z^{\prime} \in \{0, 1, \dots, N_s - 1\}$. The \ac{BS} extracts the user's received symbols from the global received frame as
\begin{align}
\label{slot_extraction}
    &\tilde{y}_{_{\widetilde{k}_\mathrm{a}}}[k^{\dagger}, l^{\dagger}] = y[k^{\dagger} + r',\; l^{\dagger} + c'], \quad \nonumber \\
& \text{for } 0 \leq k^{\dagger} < 2M_\mathrm{tile},\; 0 \leq l^{\dagger} < N_\mathrm{tile},
\end{align}
where the slot coordinates are given by $i' = \left\lfloor \frac{z'}{q} \right\rfloor, \quad
j' = \left(z'\right)_q, \,\ r' = 2M_{\text{tile}} \cdot i'$ and $c' = N_{\text{tile}} \cdot j'$. 
Next, the \ac{BS} isolates the pilot tile by selecting the subset of $\tilde{y}_{_{\widetilde{k}_\mathrm{a}}}[k^{\dagger},l^{\dagger}]$ with indices $0 \le k^{\dagger} < M_{\text{tile}}$ and $0 \le l^{\dagger} < N_{\text{tile}}$. Using the cross-ambiguity operation \cite{Ubadah_ISAC}, the estimated \ac{I/O} relationship for the $\widetilde{k}_\mathrm{a}$-th user in the singleton slot is given by
\begin{align}
\label{cross_am}
\widehat{h}_{\mathrm{eff}, \widetilde{k}_\mathrm{a}}[k^{\dagger}, l^{\dagger}]&=A_{\tilde{y}_{_{\widetilde{k}_\mathrm{a}}}, s_{_{\widetilde{k}_\mathrm{a}}}}[k^{\dagger}, l^{\dagger}] \nonumber \\
& =\sum_{k_1=0}^{M_{\text{tile}}-1} \sum_{l_1=0}^{N_{\text{tile}}-1} \tilde{y}_{_{\widetilde{k}_\mathrm{a}}}\left[k_1, l_1\right] s_{_{\mathrm{dd},\widetilde{k}_\mathrm{a}}}^*\!\left[\!k_1-k^{\dagger}\!,\!l_1-l^{\dagger}\!\right] \nonumber \\ &\cdot e^{-j 2 \pi \frac{l^{\dagger}\left(k_1-k^{\dagger}\right)}{M N}},
\end{align}
 for $0 \leq k^{\dagger} < M_{\text{tile}},\; 0 \leq l^{\dagger} < N_\text{tile}$, where $s_{_{\mathrm{dd},\widetilde{k}_\mathrm{a}}}[\tilde{k}, \tilde{l}]$ denotes the transmitted pilot signal of the $\widetilde{k}_\mathrm{a}$-th user in the tile. In \ac{DD} modulation systems such as OTFS 1.0 or Zak-OTFS (OTFS 2.0), all symbols within the frame experience nearly the same effective channel \cite{Saif_Bits2,Hadani_otfs}. Therefore, the channel estimated from the pilot tile can be reliably used to detect the data symbols in the associated data tile. 
The resulting \ac{I/O} relationship in the data tile can be represented in matrix form as
\begin{align}
    \tilde{\mathbf{y}} = {\mathbf{H}}{\mathbf{s}} + \tilde{\mathbf{n}},
\end{align}
where $\tilde{\mathbf{y}}, {\mathbf{s}}, \tilde{\mathbf{n}} \in \mathbb{C}^{M_{\text{tile}}N_{\text{tile}} \times 1}$ represent the received, transmitted, and noise vectors, respectively.  $\tilde{y}_{_{\widetilde{k}_\mathrm{a}}}[k^{\dagger},l^{\dagger}]$ and $n[k^{\dagger}, l^{\dagger}]$ correspond to the $((k^{\dagger}-M_{\text{tile}})N+l^{\dagger}+1)$-th entries of $\tilde{\mathbf{y}}$ and $\tilde{\mathbf{n }}$, respectively, for  $M_{\text{tile}}\le k^{\dagger} < 2M_{\text{tile}}$ and $0 \le l^{\dagger} < N_{\text{tile}}$. Similarly, $s_{_{\widetilde{k}_\mathrm{a}}}[\tilde{k}, \tilde{l}]$ corresponds to the $((\tilde{k}-M_{\text{tile}})N+\tilde{l}+1)$-th entry of ${\mathbf{s}}$ for  $M_{\text{tile}}\le \tilde{k} < 2M_{\text{tile}}$ and $0 \le \tilde{l} < N_{\text{tile}}$. Also, ${\mathbf{H}} \in \mathbb{C}^{M_{\text{tile}}N_{\text{tile}} \times M_{\text{tile}}N_{\text{tile}}}$ is the effective channel matrix, whose entries $H[(k^{\dagger}-M_{\text{tile}})N+l^{\dagger}+1, (\tilde{k}-M_{\text{tile}})N+\tilde{l}+1]$ are derived as in~\cite{Saif_Bits2}.  
The \ac{MMSE} equalization of the received vector $\tilde{\mathbf{y}}$ is performed as
\begin{align}
\label{mmse_op}
        \widehat{\mathbf{s}} &= \mathbf{W}_{\text{MMSE}} \, \,\tilde{\mathbf{y}}; \nonumber \\
         \mathbf{W}_{\text{MMSE}} &\Define \left( \widehat{\mathbf{H}} \widehat{\mathbf{H}}^H + \frac{1}{E_\mathrm{d}}\mathbf{R}_n \right)^{-1} \widehat{\mathbf{H}}^H,
\end{align}
where $E_\mathrm{d} = \mathbb{E}[|s_{_{\mathrm{dd},\widetilde{k}_\mathrm{a}}}[\tilde{k}, \tilde{l}]|^{2}]$ is the average energy per information symbol, $\widehat{\mathbf{H}}$ is the estimated channel matrix obtained from the effective channel response in \eqref{cross_am}, and $\mathbf{R}_n$ is the noise covariance matrix, whose entries depend on the receiver filter (see \cite{Closed_form_OTFS} for derivation).
The estimated symbol vector $\widehat{\mathbf{s}}$ is then decoded using minimum Euclidean distance detection \cite{Proakis}, yielding the estimated transmitted symbols $\breve{s}_{_{\widetilde{k}_\mathrm{a}}}[\tilde{k}, \tilde{l}]$, for $M_\mathrm{tile} \le \widetilde{k} <2M_{\text{tile}}$ and $0 \le \widetilde{l} < N_{\text{tile}}$.
Using the decoded symbols $\breve{s}_{_{\widetilde{k}_\mathrm{a}}}[\tilde{k}, \tilde{l}]$, the channel estimate $\widehat{h}_{\text{eff},\widetilde{k}_\mathrm{a}}[k, l]$ from \eqref{cross_am}, and the Zak-OTFS \ac{I/O} relationship in \eqref{rxsymbols}, the contribution of the $\widetilde{k}_\mathrm{a}$-th user to the received signal is estimated as
\begin{align}
\label{approx_rx_ka}
\widehat{y}_{_{\widetilde{k}_a}}[k^{\dagger},l^{\dagger} ] &\approx \sum_{k,l \in \mathbb{Z}} \widehat{h}_{\text{eff},\widetilde{k}_\mathrm{a}}[\!k^{\dagger}-k,l^{\dagger}-l\!] \breve{s}_{_{\text{dd},\widetilde{k}_\mathrm{a}}}[\!k,l\!] e^{j2\pi\!\left(\!\frac{(l^{\dagger} - l)k}{M}\!\right)}.
\end{align}
From the received signal model in \eqref{received_uplink} and the reconstructed signal in \eqref{approx_rx_ka}, the \ac{BS} performs \ac{SIC} by removing the contribution of the $\widetilde{k}_\mathrm{a}$-th user from all its occupied slots
\begin{align}
    \label{SIC_operation}
    &y^{(1)}[k',l'] = y[k',l']\!-\!\sum_{z'=0}^{r-1} \widehat{y}_{_{\widetilde{k}_\mathrm{a}}}[k' - 2M_{\text{tile}} i_{z'},\; l' - N_{\text{tile}} j_{z'}] 
 \nonumber \\ 
& \cdot \mathbbm{1}_{[2M_{\text{tile}} i_{z'},\; 2M_{\text{tile}}(i_{z'}+1))}(k') \cdot 
\mathbbm{1}_{[N_{\text{tile}} j_{z'},\; N_{\text{tile}}(j_{z'}+1))}(l'). 
\end{align}
After applying \eqref{SIC_operation}, the \ac{BS} re-evaluates the frame to identify newly formed singleton slots.
The decoding and \ac{SIC} procedures are repeated iteratively, each time canceling the contributions of successfully decoded users, until either all user packets have been decoded or no further singleton slots remain.
\subsection{OFDM Processing}
\label{subsec:OFDM_Processing}
In this work, we select \ac{CP}-OFDM with pilot insertion as a benchmark scheme to compare against the proposed Zak-OTFS approach. To ensure a fair comparison, the CP-OFDM system is configured with a subcarrier spacing of $\Delta f = \nu_p$, and a \ac{CP} length matched to the maximum delay spread of the channel. To enhance robustness against Doppler effects, pilot and data symbols are interleaved across alternating subcarriers. Under this configuration, a \textit{slot}, as defined consistently with Zak-OTFS, spans $M_\mathrm{sub} = 2M_{\text{tile}}$ subcarriers and $N_\mathrm{OFDM} = N_{\text{tile}}$ time indices. The resulting CP-OFDM slot structure is illustrated in Fig.~\ref{fig:Scenario_OTFSvsOFDM_SingleSlot}(a), in direct parallel to its Zak-OTFS counterpart.

To ensure parity with Zak-OTFS in terms of overall resource usage, the CP-OFDM system is configured with $N = p \cdot q \cdot N_\mathrm{OFDM}$ time indices per frame, matching Zak-OTFS in terms of time-bandwidth product. Note that this configuration follows the structure of conventional \ac{CRA} schemes based on time-slot division. Decoding at the \ac{BS} begins after the reception of all $N$ time indices, aligning with the frame-level processing strategy of Zak-OTFS. Leveraging the known preamble, the \ac{BS} first identifies singleton slots, i.e., those associated with the $\widetilde{k}_\mathrm{a}$-th active user free from collision. Channel estimation is first performed using the pilot symbols, followed by \ac{MMSE}-based interpolation. Subsequently, data detection is carried out via \ac{MMSE} equalization.

Similar to the iterative \ac{SIC} approach used in Zak-OTFS, the \ac{BS} exploits the estimated channels and the correlation structure~\cite{MIMO_OFDM} to predict the contributions of each $\widetilde{k}_\mathrm{a}$-th user's signal across other slots containing replicas.
These contributions are iteratively canceled from the received signal, progressively revealing additional singleton slots. This process continues until all user data have been decoded or no further singleton slots can be identified. In this way, \ac{CP}-OFDM follows a decoding paradigm closely aligned with that of Zak-OTFS, enabling a fair performance comparison under common assumptions. Accordingly, our focus is on evaluating each modulation scheme’s ability to support channel prediction across data carriers.

\section{Performance Evaluation}
\label{sec:Perf_Eval}

\begin{table}[!t]
\caption{Power-Delay Profile of the Veh-A Channel~\cite{Veh_A}.}
\centering
\begin{tabular}{|c!{\vrule width 0.1em}c|c|c|c|c|c|}
\hline
Path number $i$ & 1 & 2 & 3 & 4 & 5 & 6   \\
\specialrule{.1em}{.05em}{.05em} 
Rel. Delay $\tau_{i}$ $(\mu s)$ & $0$ & $0.31$ & $0.71$ & $1.09$ & $1.73$ & $2.51$   \\
\hline
Rel. Power $\frac{{\mathbb E}[ | h_{i} |^2 ]}{{\mathbb E}[ | h_{1} |^2 ]}$ (dB) & $0$ & $-1$ & $-9$ & $-10$ & $-15$ & $-20$    \\
\hline
\end{tabular}
\label{table1_veha}
\vspace{-1.3em}
\end{table}
 
\subsection{Transmission Setup}
\label{subsec:NumRes_TxSetup}
Each user encodes a $k$-bit payload using a $(n,k,t)$ binary \ac{BCH} code capable of correcting up to $t$ errors.
A portion of the $k$ bits is reserved for a \ac{CRC}, which is used to validate the decoded packet.
The resulting codeword is padded with one zero bit and mapped to a \ac{QPSK} constellation with Gray mapping.
Two BCH code configurations are used depending on the slot size: $i)$
A $(n=31,k=16,t=3)$ code for small slots ($M_\mathrm{tile}=4$, $N_\mathrm{tile}=4$);
$ii)$ A $(n=511,k=255,t=31)$ code for larger slots ($M_\mathrm{tile}=16$, $N_\mathrm{tile}=16$).
In both cases, the code rate is approximately $R_\mathrm{c} \approx 0.5$.
Furthermore, to ensure fairness in the overall use of resources between Zak-OTFS and OFDM, the OFDM subcarrier spacing is set to $\Delta f = \nu_p$ and the OFDM symbol duration to $T_\mathrm{sym} = \tau_p$, matching the \ac{DD} domain slot’s time and frequency span.
The total frame length is fixed to $N_\mathrm{s} = 128$ slots. For the Zak-OTFS-based scheme, slots are organized in a $p=8$ by $q=16$ grid in the \ac{DD} domain, yielding a square-shaped frame.


The pilot \ac{SNR} per slot is given as
\begin{equation}
    \mathrm{SNR}_\mathrm{P} = \frac{E_\mathrm{P}}{N_0 B T},
\end{equation}
where $E_\mathrm{P} = E_\mathrm{p} \cdot N_\mathrm{P}$ is the pilot energy per user per slot, $E_\mathrm{p}$ is the energy per pilot symbol, $N_\mathrm{P}$ is the number of pilot symbols per slot, $N_0$ is the noise power spectral density, $B$ is the frame bandwidth, and $T$ is the frame duration.
In Zak-OTFS, only one pilot symbol is transmitted in the pilot tile ($N_\mathrm{P} = 1$), while in OFDM, all symbols in the pilot tile are used as pilots ($N_\mathrm{P} = M_\mathrm{tile} \cdot N_\mathrm{tile}$).
Similarly, the data \ac{SNR} per slot is given as
\begin{equation}
    \mathrm{SNR}_\mathrm{D} = \frac{E_\mathrm{D}}{N_0 B T},
\end{equation}
where $E_\mathrm{D} = E_\mathrm{d} \cdot N_\mathrm{D}$ is the data energy per user per slot, $E_\mathrm{d}$ is the energy per data symbol, and $N_\mathrm{D}$ is the total number of data symbols per slot.
For both Zak-OTFS and OFDM configurations, the entire data slot region is dedicated to payload transmission, thus $N_\mathrm{D} = M_\mathrm{tile} \cdot N_\mathrm{tile}$.
The total data and pilot energy budgets are kept equal ($E_\mathrm{P} = E_\mathrm{D}$), ensuring that $\mathrm{SNR}_\mathrm{P} = \mathrm{SNR}_\mathrm{D} = \mathrm{SNR}$ for both \ac{CRA} schemes (Zak-OTFS and OFDM).
\subsection{Channel Model}
\label{subsec:NumRes_ChModel}
We adopt the six-path Veh-A channel model~\cite{Veh_A} with the power delay profile shown in Table~\ref{table1_veha}. We normalized the path energies such that $\sum_{i=1}^6 \mathbb{E}[|h_i|^2] = 1$. 
The $i$-th path Doppler shift is modeled as $\nu_i = \nu_\mathrm{max} \cos(\theta_i)$, where $\theta_i$ is an \ac{i.i.d.} random variable uniformly distributed over $[0, 2\pi)$, and $\nu_\mathrm{max} = 815$~Hz is the maximum Doppler shift assumed throughout the simulations.

\subsection{Numerical Results}
\label{subsec:Num_Res}
We organize the numerical results into three parts to evaluate the proposed uplink \ac{MMA} scheme from different perspectives:
\textit{1)} We examine an ideal interference-free scenario, where a single user transmits in a single slot without collisions for the proposed Zak-OTFS-based receiver operating in the \ac{DD} domain and the conventional OFDM-based receiver in the \ac{TF} domain (see Fig.~\ref{fig:Scenario_OTFSvsOFDM_SingleSlot}(a) and Fig.~\ref{fig:Scenario_OTFSvsOFDM_SingleSlot}(b)).
\textit{2)} We investigate the impact of channel variability on the effectiveness of \ac{SIC} for both Zak-OTFS and OFDM-based \ac{CRA} schemes (see Section \ref{sec:ProposedUplinkMA}).
\textit{3)} We present a frame-level performance analysis based on Monte Carlo simulations of the full \ac{MMA} uplink system, for both the Zak-OTFS and the conventional OFDM-based \ac{CRA} schemes.
For cases \textit{1)} and \textit{2)}, we evaluate performance in terms of \ac{PLR} versus \ac{SNR}, while for case \textit{3)}, we assess the \ac{PLR} as a function of the number of active users per frame, denoted by $K_\mathrm{a}$.
\subsubsection{Single-User Performance Analysis}
\label{subsubsec:NumRes_SingleUserPerformance}
\begin{figure}[]
    \centering
        \includegraphics[width=0.48\textwidth]{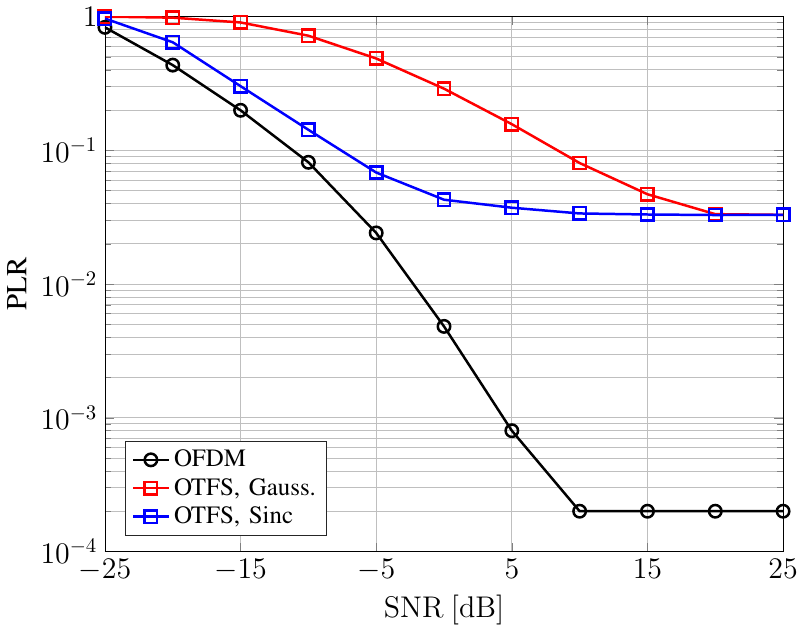}
    \caption{PLR versus SNR for a single-user, interference-free scenario with slot configuration $M_\mathrm{slot} = 8$, $N_\mathrm{slot} = 4$ and Doppler period $\nu_p = 30~\mathrm{kHz}$. The performance of Zak-OTFS with Gaussian and sinc pulse shaping is compared to conventional OFDM using a $(31,16,3)$ BCH code.}
    \label{fig:Plot_PLRvsSNR_SingleUser_OTFSvsOFDM_Mslot8_Nslot4}
\end{figure}
%
%
\begin{figure}[]
    \centering
    \includegraphics[width=0.48\textwidth]{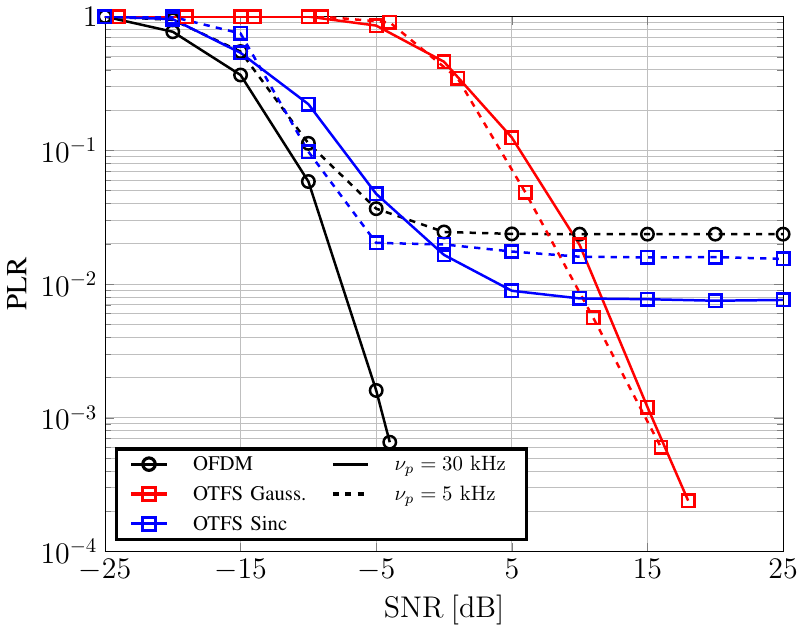}
    \caption{PLR versus SNR in a single-user, interference-free scenario with slot configuration $M_\mathrm{slot} = 32$, $N_\mathrm{slot} = 16$. Two Doppler period values are considered: $\nu_p = 30~\mathrm{kHz}$ (solid lines) and $\nu_p = 5~\mathrm{kHz}$ (dashed lines). A $(511,255,31)$ BCH code is employed. Zak-OTFS performance with Gaussian and sinc pulses is compared to conventional OFDM.}
    \label{fig:Plot_PLRvsSNR_SingleUser_OTFSvsOFDM_Mslot32_Nslot16}
\end{figure}
%
The goal is to analyze the influence of key system parameters, including the number of delay and Doppler bins per tile ($M_\mathrm{tile}, N_\mathrm{tile}$), the Doppler period $\nu_p$, and the pulse-shaping filter.


Fig.~\ref{fig:Plot_PLRvsSNR_SingleUser_OTFSvsOFDM_Mslot8_Nslot4} shows the results for a compact configuration with $M_\mathrm{tile} = N_\mathrm{tile} = 4$, leading to slot sizes $M_\mathrm{slot} = 8$ and $N_\mathrm{slot} = 4$ for Zak-OTFS, and $M_\mathrm{sub} = 8$ subcarriers and $N_\mathrm{OFDM} = 4$ OFDM symbols for the OFDM-based scheme.
The Doppler period is fixed to $\nu_p = 30~\mathrm{kHz}$, aligned with the OFDM subcarrier spacing $\Delta f$, and a $(31,16,3)$ BCH code is adopted.
In this configuration, for the Veh-A channel considered, we observe that Zak-OTFS suffers from limited resources in the \ac{DD} domain, which degrades channel estimation accuracy and results in a pronounced \ac{PLR} floor between $10^{-1}$ and $10^{-2}$ for both pulse shapes.
In contrast, OFDM benefits from a coarse subcarrier spacing, which mitigates Doppler-induced \ac{ICI} and yields superior reliability.

Fig.~\ref{fig:Plot_PLRvsSNR_SingleUser_OTFSvsOFDM_Mslot32_Nslot16} presents results for a larger configuration, with $M_\mathrm{tile} = N_\mathrm{tile} = 16$, resulting in slot sizes $M_\mathrm{slot} = 32$ and $N_\mathrm{slot} = 16$.
A $(511,255,31)$ BCH code is used, and two values of Doppler period are considered, namely, $\nu_p = 5~ \mathrm{kHz}$ and $\nu_p = 30~ \mathrm{kHz}$.
For $\nu_p = 30~\mathrm{kHz}$ (solid curves), all schemes benefit from the improved resolution. OFDM achieves extremely low \ac{PLR} for $\mathrm{SNR} \geq -5~\mathrm{dB}$.
Zak-OTFS also shows marked improvements compared to the small-slot case.
The sinc pulse shaping outperforms the Gaussian one at medium \ac{SNR}, but still exhibits an error floor around $10^{-2}$ due to residual channel estimation errors.
Conversely, the Gaussian pulse, although requiring a higher \ac{SNR} to enter the waterfall region, provides superior performance at high \ac{SNR}, thanks to its improved channel estimation capabilities.
This, however, comes at the cost of increased equalization complexity and a more gentle \ac{PLR} slope.

When the Doppler period is reduced to $\nu_p = 5~\mathrm{kHz}$ (dashed curves), OFDM performance degrades significantly.
The reduced subcarrier spacing leads to more severe \ac{ICI}, causing a \ac{PLR} floor between $10^{-1}$ and $10^{-2}$.
Zak-OTFS, by contrast, maintains stable and reliable performance for both pulse shapes, as the selected Doppler and delay period parameters ensure operation within the crystalline regime.
In particular, the conditions $\nu_p > 2\nu_\mathrm{max} = 1630~\mathrm{Hz}$ and $\tau_p = 1/\nu_p = 200~\mu\mathrm{s} > \tau_\mathrm{max} = 2.51~\mu\mathrm{s}$ guarantee that the channel interaction remains structured and predictable\footnote{$2\nu_\mathrm{max}$ and $\tau_\mathrm{max}$ denote the Doppler and delay spreads of the channel, respectively.}. This states that OFDM suffers from \ac{ICI} with a lower subcarrier spacing but the Zak-OTFS performance remains almost constant, under crystalline condition, even at a lower resolutions.

Across both scenarios, a noticeable \ac{SNR} gap emerges between Zak-OTFS and OFDM in the waterfall region, particularly when the Gaussian filter is employed.
Nonetheless, these results highlight that Zak-OTFS, especially when operated  with sufficiently high resolution, can achieve reliable performance even under severe mobility conditions.
In contrast, OFDM remains competitive in static or low-mobility environments, thanks to its efficient equalization and simplified signal processing.



\subsubsection{Impact of Channel Variability on SIC Performance}
\label{subsubsec:NumRes_SICPerformance}
\begin{figure}[]
    \centering
    \includegraphics[width=0.48\textwidth]{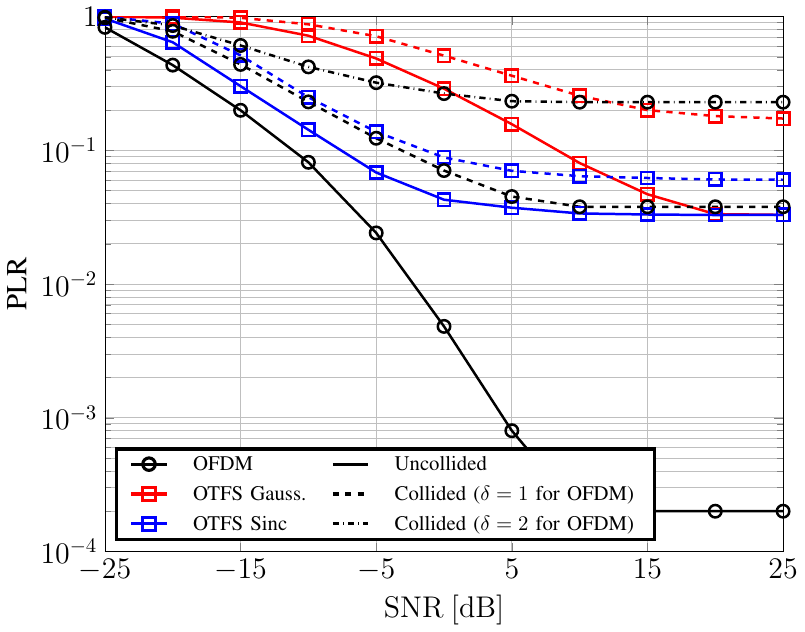}
    \caption{PLR versus SNR for a simplified two-user scenario with $M_\mathrm{slot} = 8$, $N_\mathrm{slot} = 4$ and Doppler period $\nu_p = 30~\mathrm{kHz}$. The performance of both uncollided and collided users is shown for Zak-OTFS (with Gaussian and sinc pulse shaping) and conventional OFDM.
    For OFDM, results are reported for different values of $\delta$, the temporal distance between the reference and the collided slots, highlighting the degradation in SIC performance due to channel prediction errors under high-mobility conditions.}
    \label{fig:Plot_PLRvsSNR_ToyExample_OTFSvsOFDM_Mslot8_Nslot4}
\end{figure}
\begin{figure}[]
    \centering
    \includegraphics[width=0.48\textwidth]{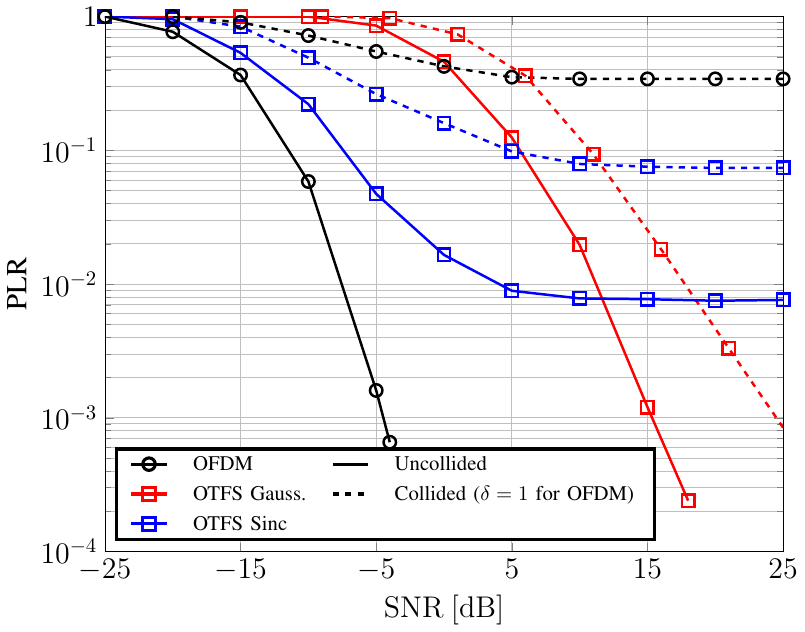}
    \caption{PLR versus SNR for a two-user scenario with slot configuration $M_\mathrm{slot} = 32$, $N_\mathrm{slot} = 16$ and Doppler period $\nu_p = 30~\mathrm{kHz}$. The PLR for collided users is shown for Zak-OTFS (using Gaussian and sinc pulse shaping) and conventional OFDM. Results highlight the severe degradation of SIC in the OFDM-based scheme, while Zak-OTFS maintains robust performance thanks to channel stability in the DD domain.}
    \label{fig:Plot_PLRvsSNR_ToyExample_OTFSvsOFDM_Mslot32_Nslot16}
\end{figure}
We now present a simplified numerical experiment to illustrate how rapidly time-varying channels affect the performance of \ac{SIC} under conventional OFDM-based processing, and how \ac{DD} domain processing using Zak-OTFS enables a more robust and effective application of \ac{SIC}.
The scenario features a transmission frame and two active users: an ``uncollided'' user and a ``collided'' user.
The uncollided user is assumed to have at least one replica transmitted without interference and at least one overlapping with other transmissions, e.g., the yellow user in Fig.~\ref{fig:Scenario_OTFSvsOFDM_Frame}(a), which has a non-interfered replica in slot~$2$ and an interfered one in slot~$6$.
In contrast, the collided user has no replica transmitted alone, and its packet can only be recovered via \ac{SIC}, after decoding and subtracting the contributions of interfering packets from the uncollided user, as shown for the blue user in Fig.~\ref{fig:Scenario_OTFSvsOFDM_Frame}(a), which is interfered in both slots~$6$ and $N_\mathrm{s} - 1$.

Focusing on the collided user, successful decoding requires effective \ac{SIC} in the slots affected by collisions.
This, in turn, demands reliable estimation of the interfering uncollided user's channel not only in the slot where it was decoded, but also in all other slots where its replicas were transmitted.
For example, if the yellow user is decoded in slot~$2$, its channel must also be estimated in slot~$6$ to enable interference cancellation.
However, due to the presence of interference, it is generally not possible to directly estimate the channel via pilot symbols in the collided slot.
As a result, the \ac{BS} must resort to \ac{MMSE}-based channel prediction, initialized from the reliable estimate obtained in the decoded slot, and attempt to extrapolate the channel evolution across time to the remaining replicas.

Under high-mobility conditions, however, conventional OFDM-based processing faces substantial challenges due to rapid channel variations across slots.
These variations degrade the accuracy of the predicted channel estimates, especially as the temporal distance between slots increases.
Consequently, the farther a replica lies from the slot where the channel was estimated, the less reliable the prediction becomes.
This significantly impairs the performance of \ac{SIC} and reduces the likelihood of successfully recovering the collided user's packet.
In contrast, Zak-OTFS-based processing in the \ac{DD} domain inherently mitigates rapid time variations.
In this domain, the channel remains approximately static across the entire frame.
Thus, once a user is successfully decoded, its channel estimate can be reliably reused in all other slots, allowing for accurate and consistent \ac{SIC} application, even in high-mobility environments.

To illustrate this effect numerically, we consider a small-slot configuration analogous to Fig.~\ref{fig:Plot_PLRvsSNR_SingleUser_OTFSvsOFDM_Mslot8_Nslot4}, with $M_\mathrm{slot} = 8$, $N_\mathrm{slot} = 4$, and Doppler period $\nu_{p} = 30~\mathrm{kHz}$.
Fig.~\ref{fig:Plot_PLRvsSNR_ToyExample_OTFSvsOFDM_Mslot8_Nslot4} reports the \ac{PLR} versus \ac{SNR} for both uncollided and collided users, comparing OFDM-based and Zak-OTFS-based \ac{CRA} schemes, with Gaussian and sinc pulse shaping filters applied to the latter.
The uncollided user performance replicate those in Fig.~\ref{fig:Plot_PLRvsSNR_SingleUser_OTFSvsOFDM_Mslot8_Nslot4}.
In the OFDM-based case, we introduce the variable $\delta$, which denotes the temporal distance, measured in slot indices, between the slot where the uncollided user's packet is decoded and the slot where its replicas must be canceled to enable decoding of the collided user (e.g., in the example mentioned above, $\delta = 4$).
Since channel prediction quality deteriorates with time, the \ac{PLR} increases as $\delta$ grows.
We report results for $\delta = 1$ (adjacent slots) and $\delta = 2$, assuming a maximum Doppler frequency of $\nu_\mathrm{max} = 815~\mathrm{Hz}$.
The results in Fig.~\ref{fig:Plot_PLRvsSNR_ToyExample_OTFSvsOFDM_Mslot8_Nslot4} reveal a clear degradation in \ac{SIC} performance with increasing $\delta$, confirming that OFDM suffers from time-selectivity.
In contrast, Zak-OTFS maintains a consistently low \ac{PLR} for the collided user, despite its slightly inferior performance for the uncollided case in this setting.

In Fig.~\ref{fig:Plot_PLRvsSNR_ToyExample_OTFSvsOFDM_Mslot32_Nslot16}, we also consider a larger-slot configuration, as in Fig.~\ref{fig:Plot_PLRvsSNR_SingleUser_OTFSvsOFDM_Mslot32_Nslot16}, with $M_\mathrm{slot} = 32$, $N_\mathrm{slot} = 16$, and $\nu_{p} = 30~\mathrm{kHz}$.
The \ac{PLR} for the collided user in this setup corresponds to the solid curves in Fig. \ref{fig:Plot_PLRvsSNR_SingleUser_OTFSvsOFDM_Mslot32_Nslot16}.
For the OFDM-based \ac{CRA} scheme, \ac{SIC} performance deteriorates dramatically even for $\delta = 1$, due to the longer slot duration.
As a result, channel prediction becomes unreliable, effectively disabling \ac{SIC} and preventing recovery of the collided packet.
Also, for Zak-OTFS with Gaussian pulse shaping, the collided user experiences only minor performance degradation compared to the uncollided case, owing to accurate channel estimation and effective interference cancellation.
On the other hand, with sinc pulses, even if the \ac{PLR} remain within acceptable bounds with an error floor between $10^{-1}$ and $10^{-2}$ the performance degrades more significantly due to its poorer estimation capabilities, which compromise the effectiveness of \ac{SIC}.

\subsubsection{Frame-Level Performance of the MMA Uplink System}
\label{subsubsec:NumRes_FramePerformance}
\begin{figure}[]
    \centering
    \includegraphics[width=0.48\textwidth]{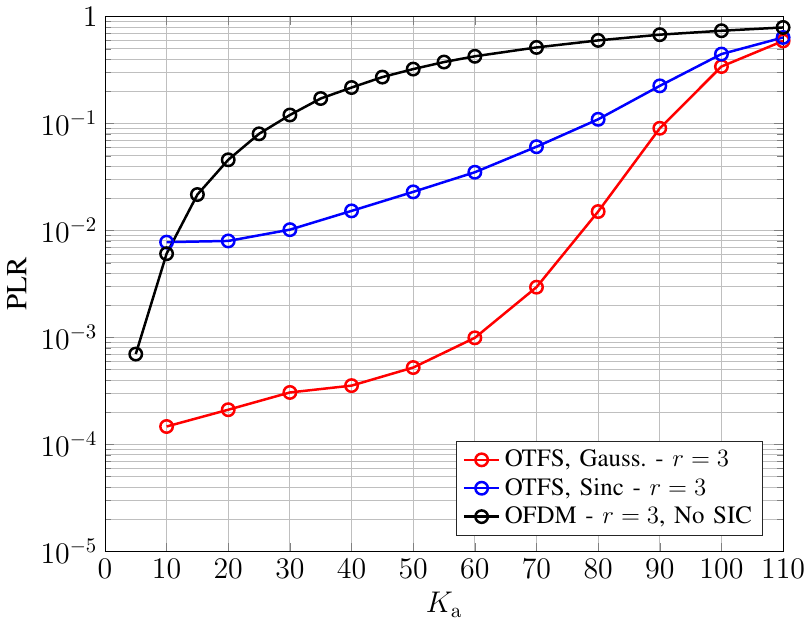}
    \caption{PLR versus number of active users per frame, $K_\mathrm{a}$, in a full-frame \ac{MMA} uplink scenario with $N_\mathrm{s} = 128$ slots, slot configuration $M_\mathrm{slot} = 32$, $N_\mathrm{slot} = 16$, and Doppler period $\nu_p = 30~\mathrm{kHz}$. Each user transmits $r = 3$ replicas of the same packet. The SNR is fixed to $25~\mathrm{dB}$. Zak-OTFS-based \ac{CRA} performance is shown for Gaussian and sinc pulse shaping, and compared against the OFDM-based scheme with no \ac{SIC}.}
    \label{fig:Plot_PLRvsKa_OTFSvsOFDM_Ns128_Mslot32_Nslot16}
\end{figure}

In this final section, we evaluate the frame-level performance of the overall \ac{MMA} uplink system under high-mobility conditions. The simulations are conducted using the larger slot configuration with $M_\mathrm{slot} = 32$, $N_\mathrm{slot} = 16$, and a Doppler period of $\nu_p = 30~\mathrm{kHz}$ at a \ac{SNR} of $25~\mathrm{dB}$.
Each user transmits $r = 3$ replicas, as often assumed in \ac{CRA} literature~\cite{haghighat2023energy, amat2018:waterfall}.
Fig.~\ref{fig:Plot_PLRvsKa_OTFSvsOFDM_Ns128_Mslot32_Nslot16} shows the \ac{PLR} as a function of the number of simultaneously active users per frame, $K_\mathrm{a}$, for both the conventional OFDM-based and the proposed Zak-OTFS-based \ac{CRA} schemes. For Zak-OTFS, results are reported for both Gaussian and sinc pulse shaping filters.

As for the OFDM-based scheme, the performance is reported under the assumption that no \ac{SIC} is applied.
As discussed above and validated via simulations, the application of \ac{SIC} is ineffective in this high-mobility regime.
The strong temporal channel variability renders inter-slot channel prediction highly inaccurate, preventing the recovery of collided packets.
Therefore, the displayed results correspond to the \ac{DSA} baseline, where performance benefits stem only from packet repetition.
However, this approach allows for low \ac{PLR} only when a small number of users is active, and rapidly saturates as $K_\mathrm{a}$ increases.

In contrast, the Zak-OTFS-based \ac{CRA} scheme exhibits strong robustness to high mobility, enabling effective \ac{SIC} and successful decoding of multiple user packets.
As a result, it achieves much better scalability.
In particular, when using the Gaussian pulse shaping filter, the Zak-OTFS system is able to support a significantly higher number of users, up to around $K_\mathrm{a} = 60$, while maintaining a \ac{PLR} below $10^{-3}$.
The observed error floor is mainly attributed to so-called \textit{unsolvable collisions}, i.e., events where two or more users fully overlap in all their replica transmissions, leaving no opportunity for successful decoding.
The Zak-OTFS scheme employing the sinc pulse filter shows both a higher error floor and a visible performance gap in the waterfall region compared to the Gaussian-filtered counterpart.
This degradation is due to the reduced accuracy of channel estimation with the sinc filter, an issue clearly observable in Fig.~\ref{fig:Plot_PLRvsSNR_ToyExample_OTFSvsOFDM_Mslot32_Nslot16}. The poorer estimation quality directly impacts \ac{SIC} effectiveness, leading to a degradation in overall performance.


\section{Conclusion}
In this work, we introduced a novel \ac{CRA} scheme operating in the \ac{DD} domain and leveraging Zak-OTFS modulation for uplink \ac{MMA}.
The proposed framework takes advantage of the predictable structure of Zak-OTFS modulation to enable effective \ac{SIC}, maintaining robust performance even in the presence of high user mobility, conditions under which conventional OFDM-based \ac{CRA} schemes suffer significant degradation.
Through extensive numerical evaluations, we demonstrated that the proposed solution offers notable gains in scalability and reliability, consistently achieving lower packet loss rates across a broad spectrum of user densities.

As part of future work, we plan to extend and integrate the proposed framework with additional components aimed at further enhancing overall system performance.
For instance, novel algorithms for multiple preamble detection should be developed to reconstruct users' repetition patterns more effectively. Moreover, advanced \ac{MPR} capabilities can be achieved through the deployment of a massive \ac{MIMO} \ac{BS}, which provides additional spatial degrees of freedom to enable user separation and channel estimation even in collided slots.
We also aim to investigate alternative unsourced random access schemes, such as coded compressed sensing and random spreading techniques beyond \ac{CRA}, and study their interplay with Zak-OTFS modulation.

This broader exploration will contribute to a more comprehensive understanding of how Zak-OTFS can influence the design and performance of future communication systems, ultimately supporting the development of more scalable, efficient, and robust next-generation wireless networks.
\bibliographystyle{IEEEtran}
\bibliography{refs}

\end{document}